\documentclass[prl,aps,showpacs,twocolumn,floatfix,superscriptaddress]{revtex4-1}
\usepackage{amsmath,amsfonts,amssymb,bm}
\usepackage{graphicx}			
\usepackage{graphics}			
\usepackage{xcolor}				
\usepackage[caption=false]{subfig}
\DeclareGraphicsExtensions{.pdf}

\newcommand{\Ref}[1]{Ref.~\cite{#1}}
\begin{document}
\title{ Pairing of $j=3/2$  fermions in half-Heusler
superconductors}

\author{P. M. R. Brydon}
\email{philip.brydon@otago.ac.nz}
\affiliation{Condensed Matter Theory Center and Joint Quantum
             Institute, Department of Physics, University of Maryland,
             College Park, MD 20742, USA}
\affiliation{Department of Physics, University of Otago, P.O. Box 56,
             Dunedin 9054, New Zealand}
\author{Limin Wang}
\affiliation{Center  for  Nanophysics  and Advanced  Materials,  Department
of Physics, University  of  Maryland,  College  Park,  MD
20742, USA}
\author{M. Weinert}
\affiliation{Department of Physics, University of Wisconsin,
                 Milwaukee, WI 53201, USA}

\author{D. F. Agterberg}
\affiliation{Department of Physics, University of Wisconsin,
                 Milwaukee, WI 53201, USA}
\begin{abstract}
  We theoretically consider the superconductivity of the
 topological half-Heusler
 semimetals YPtBi and LuPtBi. We show that pairing occurs between
 $j=3/2$ fermion states, which leads to qualitative differences from
 the conventional theory of pairing between $j=1/2$
 states. In particular, this permits Cooper pairs with quintet or septet
 total angular momentum, in addition to the usual singlet and triplet
 states. Purely on-site interactions can
 generate $s$-wave quintet time-reversal symmetry-breaking states with
 topologically nontrivial point or line nodes. These local
 $s$-wave quintet pairs reveal themselves as $d$-wave states in
 momentum space. Furthermore,
 due to the broken inversion symmetry in these materials, the $s$-wave
 singlet state can mix with a $p$-wave septet state, again with
 topologically-stable line nodes. Our analysis lays the foundation for
 understanding the unconventional superconductivity of the half-Heuslers.
\end{abstract}
\date{\today}
\maketitle

The concept of topological order is now firmly established as a
key characteristic of condensed matter systems.
Although fundamentally different from
spontaneous symmetry-breaking order, there is 
much interest in whether a nontrivial relationship between the two
exists. 
A materials class in which to systematically
explore this interplay are the
ternary half-Heusler compounds, in particular {\it R}PtBi and {\it
R}PdBi, where {\it R} is a rare earth. 
Many
of these systems are predicted to show an inversion between the
$p$-orbital-derived $j=3/2$ $\Gamma_8$ and the $s$-orbital-derived
$j=1/2$ $\Gamma_6$ bands~\cite{hhtop}, a precondition for a
topological insulator state.
These half-Heuslers also display symmetry-broken
ground states: Most are either
antiferromagnetic~\cite{can91,nak15} or
superconducting~\cite{gol08,but11,taf13,xu14}, or show
a coexistence of the two~\cite{pan13,nit15,nak15}.
Excitingly, there
is now compelling evidence that the superconductivity of YPtBi is
unconventional: Upper critical field measurements are inconsistent
with singlet pairing~\cite{bay12}, while the low-temperature
penetration depth indicates the presence of line nodes~\cite{jp15}. A surface
nodal superconducting state in LuPtBi with a $T_c$ significantly
higher than in the bulk has also been reported~\cite{ak15}.

The band inversion predicted for YPtBi and LuPtBi
implies a fundamental difference from
most other superconductors: In these materials, the chemical
potential lies 
close to the four-fold degeneracy point of the $\Gamma_8$ band,  and
a 
microscopic theory of the superconductivity must therefore
describe the pairing between $j=3/2$
fermions. This is highly unusual, since the
four-fold degeneracy is typically
split by crystal fields and spin-orbit interactions to the
 two-fold degeneracy dictated by parity and time-reversal
 symmetries,  yielding the conventional pseudospin-$1/2$
 description of Cooper pairing.

In this Letter we investigate the possible superconducting
states of YPtBi and LuPtBi. Our starting point is a generic
${\bf k}\cdot{\bf p}$ model for the low-energy states of the $\Gamma_8$
band, which qualitatively captures the {\it ab initio} band
structure. Although both symmetric (SSOC) and
antisymmetric spin-orbit coupling (ASOC) lift the four-fold
degeneracy away from the $\Gamma$ point, the electronic states
nevertheless maintain their $j=3/2$ character. This has  important
consequences for the superconductivity. In particular, there are six
distinct  on-site pairing states: one
corresponds to the conventional $J=0$ singlet solution, while the other
five are $J=2$ {\it quintet} states.  Pairing in the
latter channels generically leads to nodal time-reversal
symmetry-breaking (TRSB) states, but strongly depends upon
the SSOC. Due to the absence of centrosymmetry, the on-site singlet
solution can mix with a $p$-wave $J=3$ {\it septet}
state, potentially yielding a nodal gap which is insensitive to
the pair-breaking effect of the ASOC.
The essential
role of spin-orbit coupling in selecting the pairing state has been
overlooked in previous works~\cite{majorana}, which
examined pairing in $j=3/2$ bands in the context of realizing
topological surface states. Such considerations also do not
arise in the pairing
of spin-$3/2$ particles in cold atomic
gases~\cite{ho99}. Our work therefore lays the foundation for
understanding the superconductivity of topological half-Heusler compounds.

\begin{figure}%
    \includegraphics[width=\columnwidth]{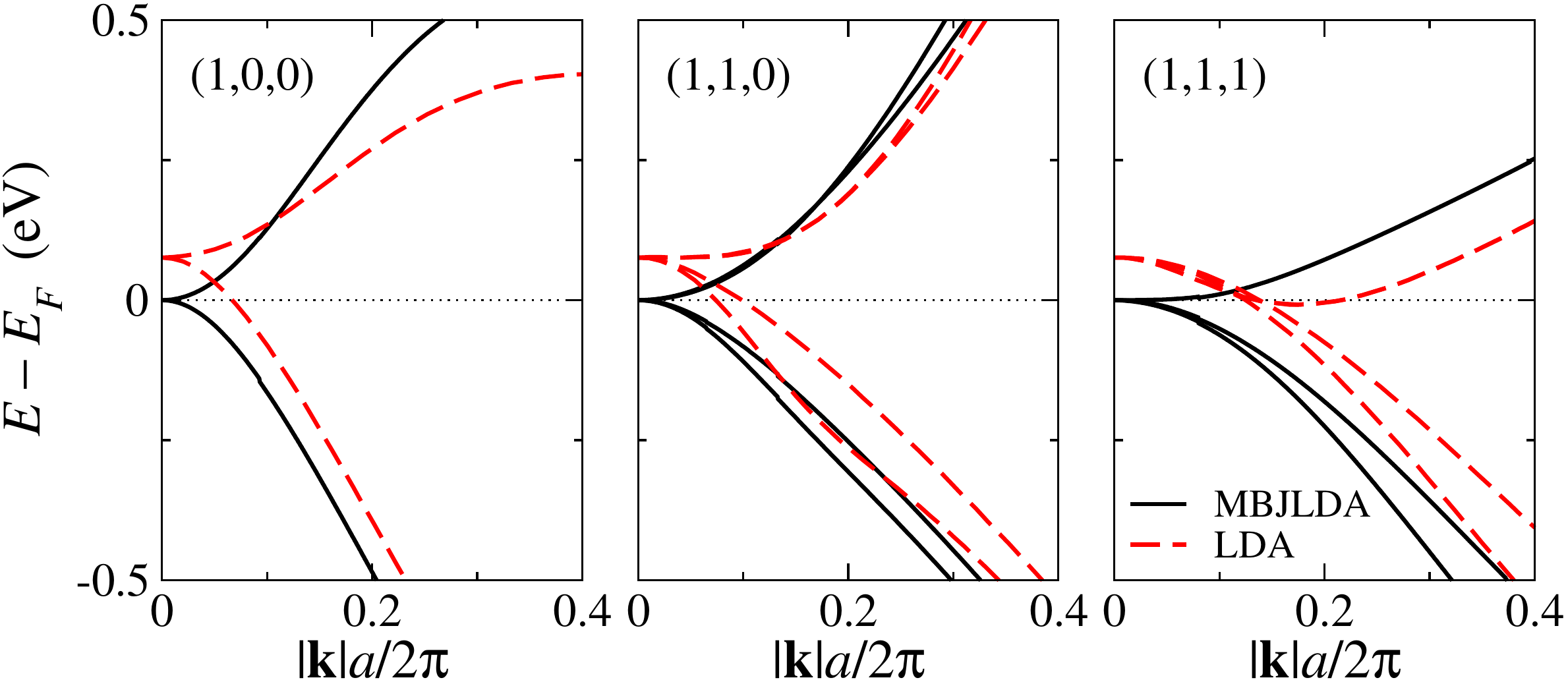}
    \caption{(Color online) Comparison of MBJLDA and LDA results for
    the $\Gamma_8$ band of YPtBi along high symmetry directions close
    to the $\Gamma$ point. The dotted line indicates the Fermi energy
    and $a$ is the lattice constant.}
    \label{abinitio}
\end{figure}

{\it Generic ${\bf k}\cdot{\bf p}$ model for half-Heusler
semimetals}---Band structure calculations for YPtBi and LuPtBi
indicate that the electronic states near the chemical potential
arise from the $j=3/2$ $\Gamma_8$
representation, where the $j=3/2$ total angular momentum is due to the
spin-orbit coupling of spin $s=1/2$ electrons in $l=1$ $p$-orbitals
of Bi.
 In
Fig.~\ref{abinitio} we compare {\it ab initio} predictions for the
$\Gamma_8$ band in YPtBi. We note that the band
structure calculated using different exchange correlation potentials
differ to some degree \cite{hhtop,suppl}. In particular, whereas the
local-density approximation (LDA) predicts a compensated
semimetal,
the modified Becke and
Johnson potential (MBJLDA) yields a zero band-gap semiconductor.
The two schemes are in much better agreement for
LuPtBi~\cite{suppl}.
Further details of the {\it ab initio} calculations,
including hybrid HSE06 functional results confirming the band inversion, are
given in the supplemental material~\cite{suppl}.
In either case it
is possible to model the band structure near the $\Gamma$ point with
a ${\bf k}\cdot {\bf p}$ theory. Such a theory was originally discussed
by Dresselhaus \cite{dre55}; up to quadratic order in $k$ the
single-particle Hamiltonian is
\begin{eqnarray}
H&=&\alpha k^2 +\beta\sum_i k_i^2\check{J}_i^2 {+\gamma \sum_{i\ne
j}k_ik_j\check{J}_i\check{J}_j} \notag \\&&+
\delta \sum_i k_i(\check{J}_{i+1}\check{J}_i\check{J}_{i+1}-\check{J}_{i+2}\check{J}_i\check{J}_{i+2})
\label{Ham}
\end{eqnarray}
 where $i={x,y,z}$ and $i+1=y$ if $i=x$, {\it etc}., and $\check{J}_i$
are $4\times 4$ matrices corresponding to the angular
momentum operators for $j=3/2$. The first line of Eq.~(\ref{Ham}) is
 the Luttinger-Kohn model, which is invariant under inversion 
and involves SSOC terms proportional
to $\beta$ and $\gamma$. The second line is odd
under inversion 
and generalizes the ASOC 
discussed in the context of $j=1/2$
noncentrosymmetric superconductors \cite{bau12}. Although this
model qualitatively captures the predicted band structure, it is
necessary to include higher-order terms in the ${\bf k}\cdot{\bf p}$
expansion to achieve quantitative agreement~\cite{suppl}. Since
including these additional terms does not alter our conclusions about
the
superconductivity, but significantly complicates the analysis,
we neglect them in the following.

Even with this simplification, it is not generally possible
to analytically diagonalize the
Hamiltonian~(\ref{Ham}).
For our study of the superconductivity, however, we only require an
effective low-energy model valid close to the Fermi surface. We obtain
this by treating the ASOC as a perturbation of the Luttinger-Kohn
bands, which
is justified when the
characteristic ASOC energy $\sim\delta k_F$ is small
compared to the chemical potential measured from the four-fold
degeneracy point. Experiments showing a low density of hole
carriers~\cite{but11,bay12}, and the predicted very
weak ASOC splitting, are consistent with this condition.

\begin{figure}
\includegraphics[width=0.52\columnwidth]{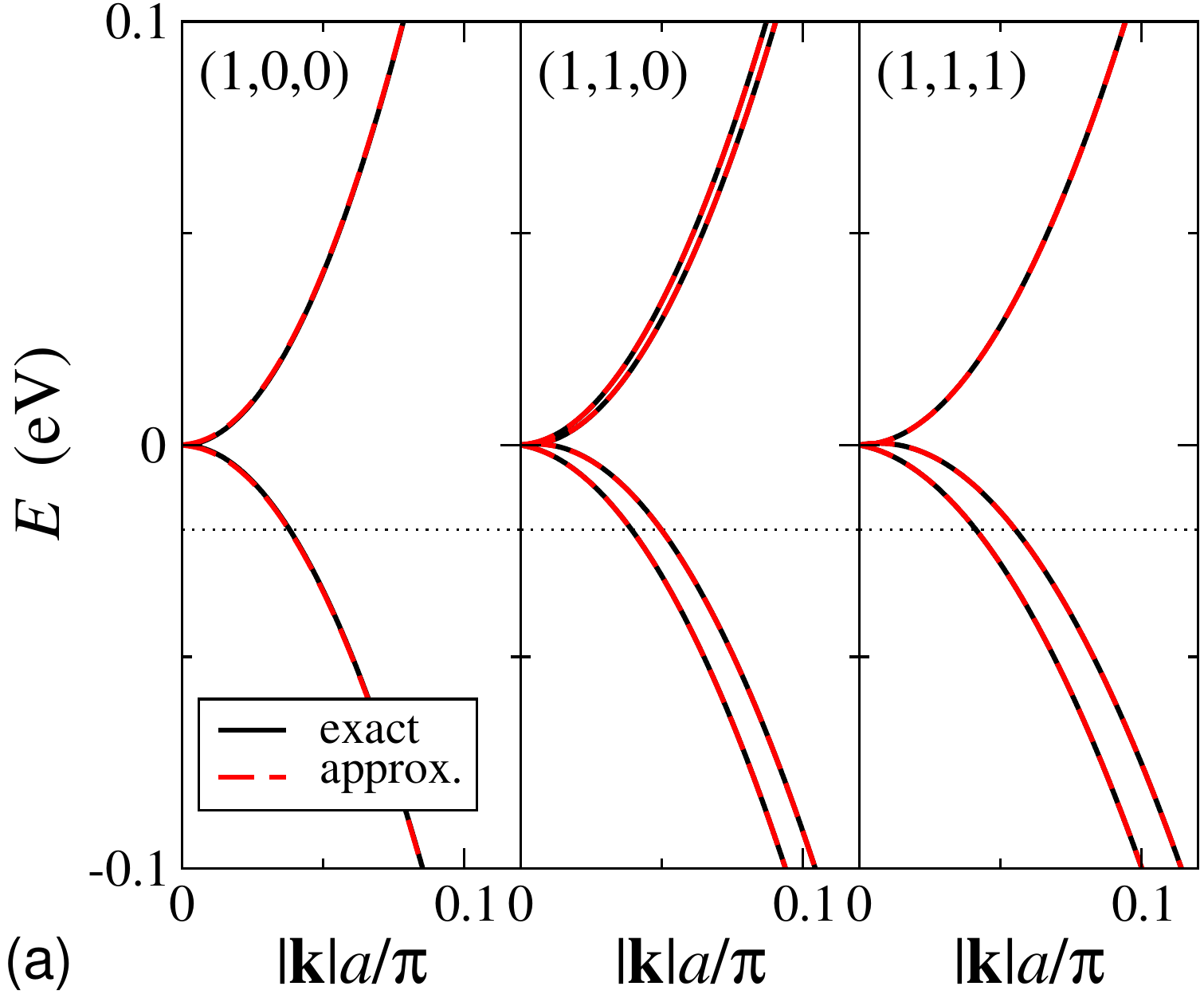}\hspace*{0.02\columnwidth}\includegraphics[width=0.46\columnwidth]{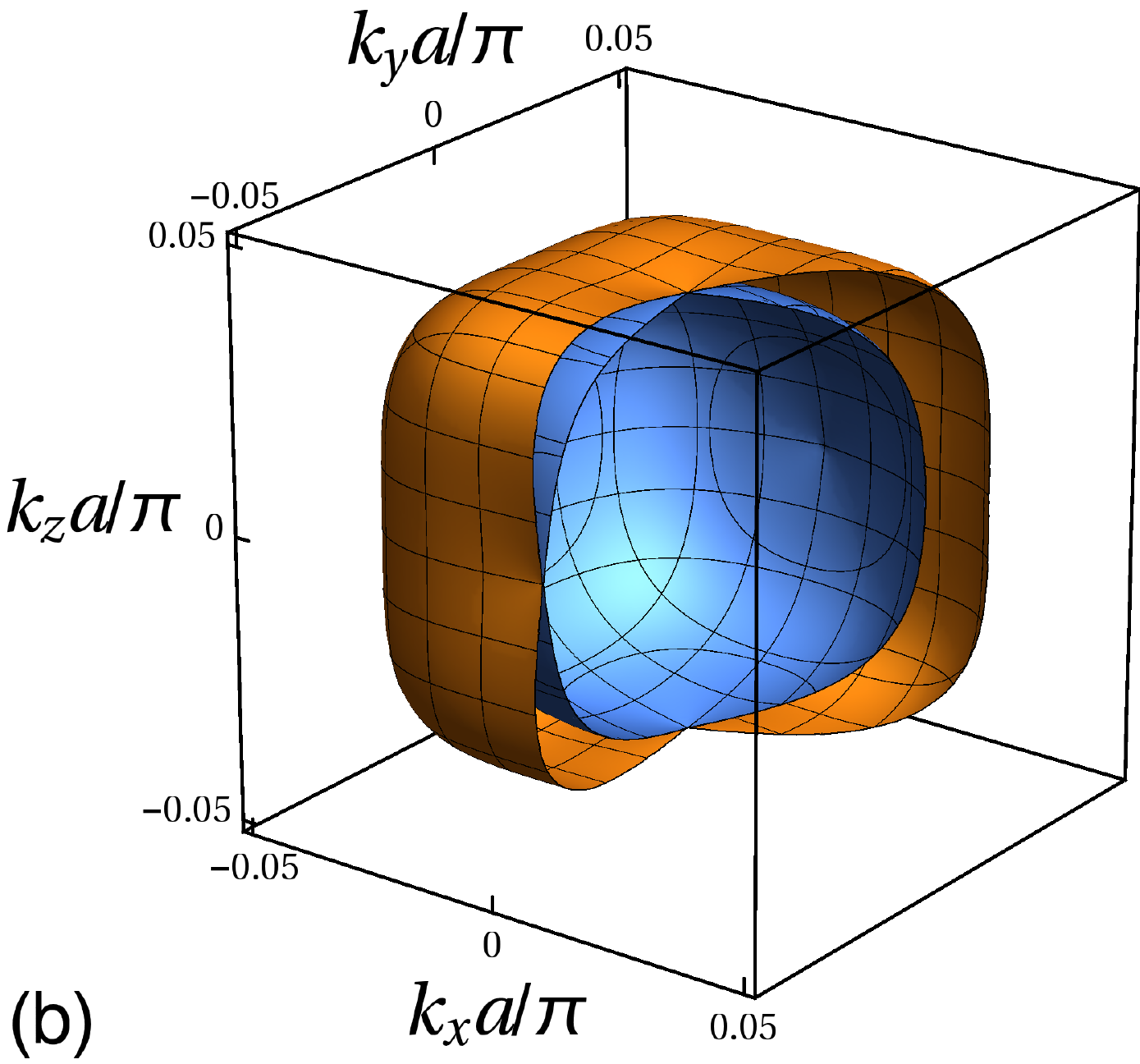}
\caption{(Color online) (a) Comparison of exact and
approximate small-ASOC dispersions along high symmetry directions. (b)
Cut-away of spin-orbit-split holelike
Fermi surfaces for $\mu=-20$meV. In all figures we take the
parameters of the ${\bf k}\cdot{\bf p}$ Hamiltonian~(\ref{Ham}) to be $\alpha=20
(a/\pi)^2$eV, $\beta = -15 (a/\pi)^2$eV, $\gamma=-10 (a/\pi)^2$eV, and
$\delta = 0.1(a/\pi)$eV. }
\label{kpHam}
\end{figure}

The eigenstates of
the Luttinger-Kohn model
are doubly degenerate and can be
labelled by pseudospin-$1/2$ indices.
The dispersions are given by
\begin{equation}
\epsilon_{\bf k,\pm}=\left(\alpha
+{ \frac{5}{4}}\beta\right)|{\bf k}|^2\pm\beta\sqrt{\sum_{i}\left[k_i^4+\left(\tfrac{3\gamma^2}{\beta^2}-1\right)k_i^2k_{i+1}^2\right]}. \label{eq:centrodisp}
\end{equation}
We now include the ASOC as a first-order perturbation by projecting
the ASOC into the pseudospin basis
for each band. We hence obtain two effective pseudospin-$1/2$ Hamiltonians
\begin{equation}
H_{\rm{eff},\pm}=\mathcal{P}_{\pm}\mathcal{U}^{\dagger}H\mathcal{U}\mathcal{P}_{\pm}=\epsilon_{\bf
k,\pm}\hat{s}_0 +{\bf g}_{{\bf k},\pm}\cdot \hat{\bf s} \label{eq:effHam}
\end{equation}
where $\mathcal{P}_{\pm}$ projects into the pseudospin states of the
 $\epsilon_{{\bf k},\pm}$ bands~(\ref{eq:centrodisp}),
$\mathcal{U}$ is the unitary operator that
diagonalizes $H$ with the ASOC set to zero, and $\hat{s}_\mu$ are the
Pauli matrices for the pseudospin. The vector ${\bf g}_{{\bf
k},\pm} = -{\bf g}_{-{\bf k},\pm}$ represents the effective ASOC in the
pseudospin-$1/2$ basis of the band $\epsilon_{{\bf k},\pm}$. While
the {\it orientation} of ${\bf g}_{{\bf k},\pm}$ depends on the arbitrary
choice of pseudospin basis, the
{\it magnitude} of ${\bf g}_{{\bf k},\pm}$ is independent of this
choice and can be written
\begin{widetext}
\begin{equation}
|{\bf g}_{\bf
 k,\pm}|^2=\frac{9\delta^2}{16} \left(\frac{\sum_i[(1+\frac{4\gamma^2}{\beta^2})k_i^4(k_{i+1}^2+k_{i+2}^2)+(\frac{4\gamma^2}{\beta^2}-2)k_i^2k_{i+1}^2k_{i+2}^2]}{\sum_i\left[k_i^4+(\frac{3\gamma^2}{\beta^2}-1)k_i^2k_{i+1}^2\right]} \pm\frac{4\frac{\gamma}{\beta}\sum_ik_i^2k_{i+1}^2}{\sqrt{\sum_i\left[k_i^4+(\frac{3\gamma^2}{\beta^2}-1)k_i^2k_{i+1}^2\right]}}\right)
\end{equation}
\end{widetext}
Note that along the $(1,1,1)$ direction this becomes 
$|{\bf
g}_{{\bf k},\pm}|^2=\frac{3}{4} \delta^2
k^2[1\pm \rm{sgn}(\gamma/\beta)]$,  which is
vanishing in one band but nonzero in the other. In the usual $j=1/2$
case, however, symmetry dictates that the
ASOC must vanish along this direction~\cite{bau12};  the
spin-orbit splitting of
one of the bands therefore reflects the presence of $j=3/2$ physics even
in our effective
pseudospin-$1/2$ description.
The effective Hamiltonians~(\ref{eq:effHam}) can be readily
diagonalized and yield
the dispersions $E_{{\bf k},\eta=\pm,\nu=\pm} = \epsilon_{{\bf k},\eta} + \nu|{\bf
g}_{{\bf k},\eta}|$,
where the values of $\eta$ and $\nu$ are independent of one
another.  As shown in Fig.~\ref{kpHam}(a), this approximate
dispersion is in excellent agreement with the full numerical solution
of the ${\bf k}\cdot{\bf p}$ Hamiltonian, and yields typical spin-orbit
split holelike Fermi surfaces plotted in~Fig.~\ref{kpHam}(b).

{\it Superconductivity}---In the conventional theory of
superconductivity, a Cooper pair
constructed from two $j=1/2$ fermions has either 
total angular momentum $J=0$ (singlet) or $J=1$ (triplet),
which by fermion antisymmetry correspond to even- and odd-parity
orbital states,
respectively. For the pairing of the $j=3/2$ states 
in the half-Heuslers, however, we must additionally allow for 
 $J=2$ (quintet) and $J=3$ (septet) pairing, again
corresponding to even- and odd-parity orbital
wavefunctions. These extra pairing channels already
manifest themselves in an expanded variety of on-site ($s$-wave) pairing:
While there is a
single $J=0$ state, there are five distinct types of on-site Cooper
pair with $J=2$. The six local Cooper pair operators
$b_{l,i}=\sum_{m,m'}\Gamma^l_{m,m'}c_{m,i}c_{m',i}$  are defined and
classified according to the tetrahedral point group symmetry in
Table~\ref{Cooper}.

\begin{table}
\begin{tabular}{|c|c|c|}
   \hline
   Representation & Cooper Pair & $J$\\
   \hline
   $A_1$ &  $c_{3/2}c_{-3/2}-c_{1/2}c_{-1/2}$ & singlet\\
\hline
$E$ & $c_{3/2}c_{-3/2}+c_{1/2}c_{-1/2}$ & quintet\\
& $c_{3/2}c_{1/2}+c_{-1/2}c_{-3/2}$ & quintet\\
\hline
$T_2$ & $c_{3/2}c_{-1/2}+c_{1/2}c_{-3/2}$ & quintet\\
& $-i\left(c_{3/2}c_{-1/2}-c_{1/2}c_{-3/2}\right)$ & quintet\\
& $-i\left(c_{3/2}c_{1/2}-c_{-1/2}c_{-3/2}\right)$ & quintet\\
   \hline
 \end{tabular}
\caption{On-site Cooper pair operators for $j=3/2$ pairing. The first
    column gives the representation of $T_d$,
the second shows the form of the local Cooper pair
 operator (with site index suppressed), and the last column gives the
 total angular momentum state.}
\label{Cooper}
\end{table}

In terms of these  basis functions, the on-site pairing
interaction will have the form $H_{\text{pair}}=\sum_l V_l
b_{l,i}^{\dagger} b_{l,i}$
with one  potential $V_l$ for each tetrahedral representation.
Treating this
within a usual mean-field theory yields a pairing term of
the form
\begin{equation}
H_{\text{pair}}=\sum_{\bf k}\sum_{j,j'=-3/2}^{3/2}\Big\{\Delta_{j,j'}
c^{\dagger}_{{\bf k},j}c^{\dagger}_{-{\bf k},j'}+\text{H.c.}\Big\}.
\end{equation}
It is instructive to project the $\Delta_{j,j'}$
into the pseudospin basis of the
$\epsilon_{{\bf k},\pm}$ bands,
$\Delta_{\text{eff},\pm}({\bf
k})=\mathcal{P}_{\pm}\mathcal{U}^{\dagger}\Delta\mathcal{U}^*\mathcal{P}_{\pm}$.  In
all cases the even parity of the pairing yields a pseudospin-singlet
gap. Neglecting higher-order corrections, for on-site Cooper
pairs in representation $A_1$, we find
\begin{equation}
\Delta^{A_1}_{\text{eff},\pm}=\Delta_s i\hat{s}_y,
\end{equation}
for on-site $E$ Cooper pairs we find
\begin{equation}
\Delta^{E}_{\text{eff},\pm}=\pm\frac{\beta}{4}\frac{\eta_1(2
k_z^2-k_x^2-k_y^2)+\eta_2\sqrt{3}(k_x^2-k_y^2)}{\sqrt{\beta^2\sum_ik_i^4+(3\gamma^2-\beta^2)\sum_ik_i^2k_{i+1}^2}}i\hat{s}_y,
\end{equation}
where ${\bm \eta}=(\eta_1,\eta_2)$ is a two-component order parameter,
and for on-site $T_2$ Cooper pairs we find
\begin{equation}
\Delta^{T_2}_{\text{eff},\pm}=\pm\frac{\sqrt{3}\gamma}{2}\frac{l_1k_yk_z+l_2k_xk_z+l_3k_xk_y}{\sqrt{\beta^2\sum_ik_i^4+(3\gamma^2-\beta^2)\sum_ik_i^2k_{i+1}^2}}i\hat{s}_y,
\end{equation}
which is characterized by the three-component order parameter
$\bm{l}=(l_1,l_2,l_3)$.  The effective gaps of the
quintet pairing states have $d$-wave form factors, which
reflects the $J=2$ total angular momentum of the Cooper pairs.
The $d$-wave symmetry is therefore a robust result, and does not
depend on the specific parameters of our ${\bf k}\cdot {\bf p}$ Hamiltonian.
Before discussing each of these cases in detail, we
 note an important property of the $E$ and $T_2$ states:
The effective gaps,
and therefore $T_c$, depend strongly on the SSOC terms
in~Eq.~(\ref{Ham}). Specifically, the effective gap for the $E$
states is vanishing unless $\beta\neq0$, while the $T_2$ states only
open a gap at the Fermi surface if $\gamma\neq0$. Consequently, a
spatial variation of the spin-orbit coupling (as might appear near surfaces or
interfaces) can dramatically
change $T_c$ for these solutions. We speculate that this may explain
the enhanced $T_c$ observed at the surface of
LuPtBi~\cite{ak15}.

\begin{figure}
\begin{center}
\includegraphics[width=\columnwidth]{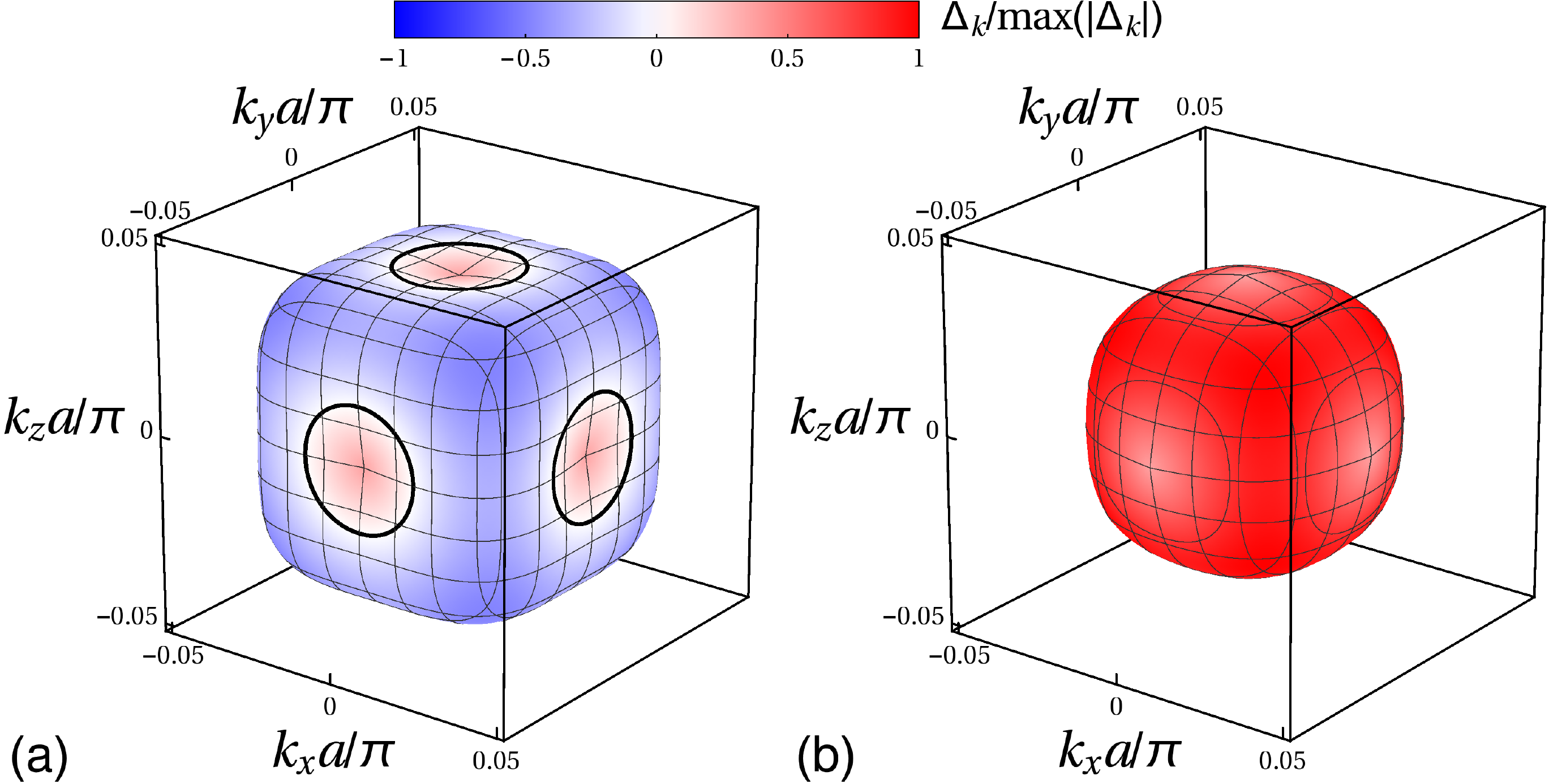}
\end{center}
\caption{(Color online) Typical mixed singlet-septet $A_1$ pairing
state with (a) a nodal gap on the larger Fermi surface and (b) a full
gap on the smaller Fermi surface. }
\label{fig:mixed}
\end{figure}

{\it The $A_{1}$ pairing state}---The on-site
pairing in the $A_1$ channel corresponds to the conventional isotropic
$s$-wave singlet state. It is therefore interesting
to consider the effect of the broken inversion symmetry, which in
$j=1/2$ noncentrosymmetric superconductors generates a
mixed-parity state with both singlet and triplet pairing~\cite{fri04}.
For $T_d$ symmetry, the lowest orbital-angular-momentum $A_1$ triplet
state is $f$-wave, which for small $k$ gives gap functions on the two
spin-split  $j=1/2$
Fermi surfaces 
$\Delta({\bf
k})=\Delta_s\pm \Delta_f\sqrt{\sum_{i}k_i^2(k_{i+1}^2-k_{i+2}^2)^2}$.
This state exhibits
line nodes if
the $f$-wave triplet gap $\Delta_f$ is larger than the $s$-wave
singlet gap $\Delta_s$. However, dominant $f$-wave
symmetry of the Cooper pairs is highly unlikely
if quasi-local interactions give rise to
superconductivity~\cite{kon89}; such interactions would more
plausibly give rise to a $p$-wave state. For the $j=3/2$
case considered here, however, a $p$-wave state with $A_1$ symmetry
exists: In the basis $(c_{{\bf k},3/2},c_{{\bf
k},1/2},c_{{\bf k},-1/2},c_{{\bf k},-3/2})$ it has gap function
\begin{equation}
\Delta({\bf k})=\Delta_p\left(
                  \begin{array}{cccc}
                    \frac{3}{4}k_- & \frac{\sqrt{3}}{2}k_z &
                  \frac{\sqrt{3}}{4}k_+ & 0 \\
                    \frac{\sqrt{3}}{2}k_z & \frac{3}{4}k_+ & 0 & -\frac{\sqrt{3}}{4}k_- \\
                    \frac{\sqrt{3}}{4}k_+ & 0 & -\frac{3}{4}k_- & \frac{\sqrt{3}}{2}k_z \\
                    0 & -\frac{\sqrt{3}}{4}k_- & \frac{\sqrt{3}}{2}k_z & -\frac{3}{4}k_+ \\
                  \end{array}
                \right) \label{eq:septet}
\end{equation}
where $k_{\pm}=k_x\pm i k_y$. This constitutes a {\it septet} pairing
state with total $J=3$.
 Projecting the gap into the effective pseudospin-$1/2$
bands, we find that $\Delta_{\text{eff},\pm}=({\bf d}_{{\bf
k},\pm}\cdot\hat{\bf s})i\hat{s}_y=(\Delta_p/\delta)({\bf g}_{{\bf
k},\pm}\cdot\hat{\bf s})i\hat{s}_y$, i.e. the ${\bf
d}$-vector of the effective pseudospin-triplet state is parallel to
the effective ASOC vector ${\bf g}_{{\bf k},\pm}$. As pointed
out in~\Ref{fri04}, 
this alignment makes the
gap $\Delta_{\text{eff},\pm}$ immune to
the pair-breaking effect of the ASOC; for sufficiently large
ASOC, it is the only stable
odd-parity gap. Importantly, when mixed with a subdominant $s$-wave
singlet state, the resulting gap displays line nodes on one of the spin-split
Fermi surfaces, as shown
in~Fig.~\ref{fig:mixed}. These nodes are topologically protected and
lead to zero-energy flat band surface
states~\cite{sch15}.

\begin{figure}
\begin{center}
\includegraphics[width=\columnwidth]{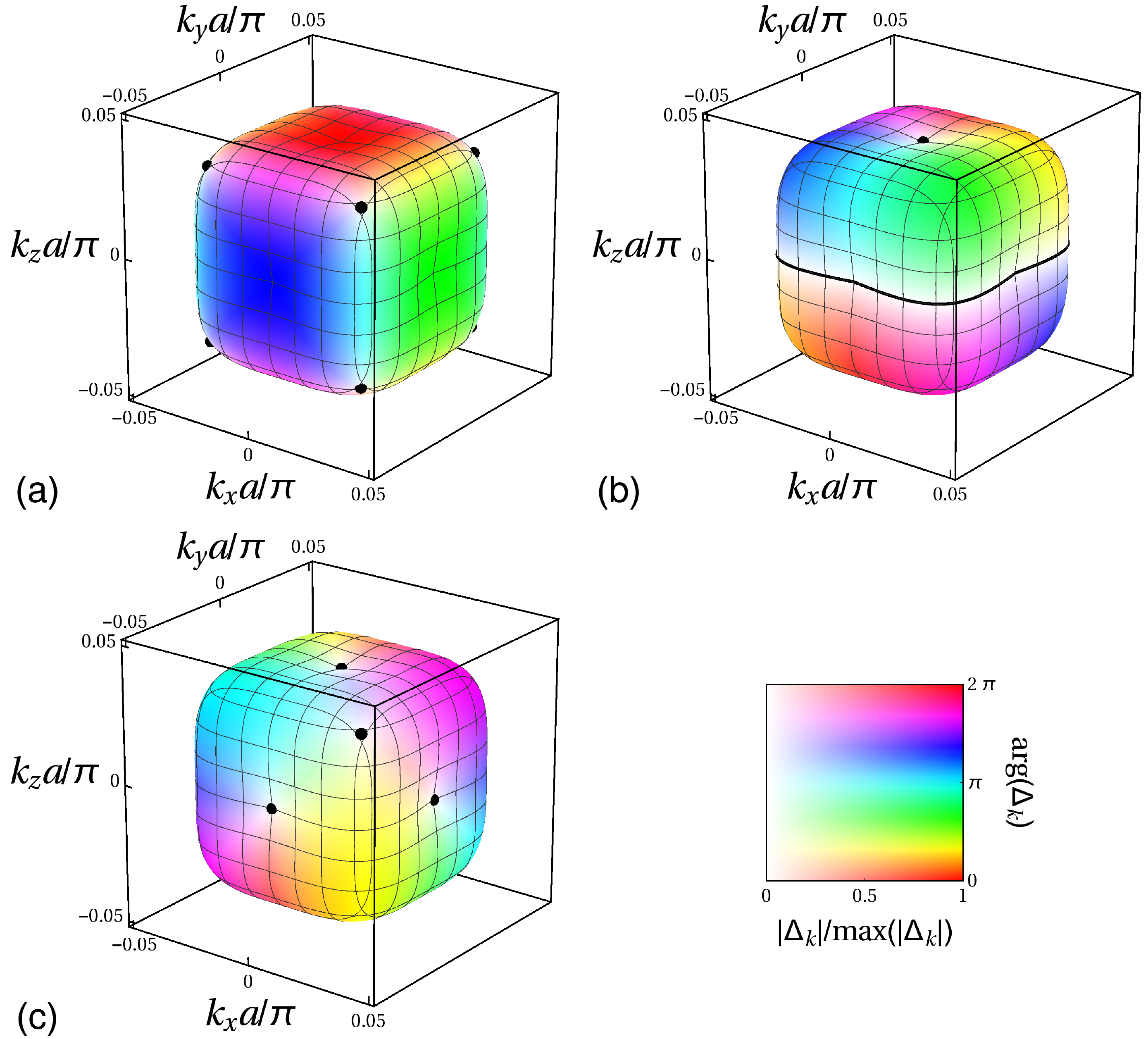}
\end{center}
\caption{(Color online) Time-reversal symmetry-breaking quintet
pairing states: (a) the $E$ pairing state; (b) the $T_2$ pairing state
with ${\bf l}=(1,i,0)$; (c) the $T_2$ pairing state with ${\bf
l}=(1,-e^{2\pi i/3},e^{4\pi i/3})$. The color indicates the phase while
the saturation gives the
gap magnitude. Black points or lines indicate nodes of the gap.}
\label{fig:trsb}
\end{figure}

{\it The $E$ pairing state}---The properties of the $E$
superconducting state depends upon the two-dimensional order parameter
${\bm \eta}=(\eta_1,\eta_2)$. The free energy expansion for the
$E$ pairing state in point group $T_d$ is the same as that for an
$E_g$ state in point group $O_h$~\cite{sig91}, from which we deduce
$f_E=\alpha{\bm \eta}\cdot{\bm \eta}^*+\beta_1({\bm \eta}\cdot{\bm \eta}^*)^2+\beta_2(\eta_1\eta_2^*-\eta_2\eta_1^*)^2$.
In general there are three ground states: ${\bm \eta}=(1,0)$, $(0,1)$, and
$(1,i)$.
In the weak-coupling limit we find $\beta_1=3\beta_2>0$ {\it independent} of the
particular form of the gap basis functions or the shape of the Fermi
surface, ensuring that the TRSB state
${\bm \eta} = (1,i)$ is most stable. The effective gap, shown
in~Fig.~\ref{fig:trsb}(a), has topologically-protected Weyl point nodes
that generate arc
surface states~\cite{sch15}. Although point nodes at first seem
inconsistent with the observation of line nodes,
it is possible that a
point node state with impurities resembles a clean line node
state \cite{nag15,hir88}, and hence it cannot be excluded
as a possible pairing state in YPtBi.

{\it The $T_2$ pairing state}---The gap function for $T_2$
pairing is controlled by the three-dimensional order parameter
$\bm{l}=(l_1,l_2,l_3)$. Similar to the $E$ pairing state,
the free energy expansion for the $T_2$ pairing in the $T_d$ point
group is identical to that for $T_{2g}$ pairing in the point group
$O_h$~\cite{sig91},
i.e. $f_{T_2}=\alpha\bm{l}\cdot \bm{l}^*+\beta_1(\bm{l}\cdot\bm{l}^*)^2
+\beta_2|\bm{l}\cdot\bm{l}|^2+\beta_3(|l_1|^2|l_2|^2+|l_1|^2|l_3|^2+|l_3|^2|l_2|^2)$. This admits four distinct ground states: ${\bm l}=(1,0,0)$,
$(1,1,1)$, $(1,e^{2\pi i /3},e^{4\pi i/3})$, or
$(1,i,0)$. Again assuming weak coupling, the
parameters in the free energy expansion satisfy $\beta_1>0$,
$\beta_2>0$, and
$\beta_3=2\beta_2-\beta_1$, which implies that one of the two TRSB
states is realized. The particular state
depends on the detailed  form of the gap basis functions and the shape
of the Fermi surface. We plot the corresponding effective gaps
in~Fig.~\ref{fig:trsb}(b) and (c). Both
these gaps have interesting topological properties and surface
states~\cite{sch15}. Given that line nodes have been observed, the ${\bm
l}=(1,i,0)$ solution is of particular interest.

 {\it Conclusions}---In this Letter we have investigated
possible pairing states of the unconventional noncentrosymmetric
superconductors YPtBi
and LuPtBi. The inverted band structures of these topological
semimetals implies pairing of $j=3/2$ fermions,
permitting Cooper pairs in a quintet or septet total angular
momentum state. On-site quintet pairing generically leads
to nodal TRSB superconducting states, which could be detected by
magneto-optical
Kerr effect or $\mu$SR measurements. 
Alternatively, a nodal
time-reversal symmetric gap can arise from the admixture
of a $p$-wave septet state with an on-site singlet state. Spin-orbit
coupling strongly influences the stability of these states. The
similar electronic structure of the topological half-Heusler
compounds makes our analysis relevant to the superconductivity of
the entire materials class. Although we have not considered a pairing
mechanism, the low carrier
density makes a conventional Eliashberg theory unlikely~\cite{mei15}.
We note that pairing of $j=3/2$ fermions is not necessarily limited to
the half-Heuslers: 
the four-fold degeneracy of the $\Gamma_8$ bands also occurs in
materials
with $O$, $T$, and $O_h$ point group symmetries, permitting the
exotic superconducting states discussed here.

\begin{acknowledgments}
We acknowledge support from Microsoft Station Q,
LPS-CMTC, and JQI-NSF-PFC (P.M.R.B), J. Paglione and the
U.S. Department of Energy Early Career award DE-SC-0010605 (L.W.), and
the NSF via DMREF-1335215 (D.F.A. and M.W.). The authors thank
A. Kapitulnik, H. Kim, and J. Paglione for sharing unpublished
experimental data and for stimulating discussions. C. Timm is thanked for
helpful comments on the manuscript.
\end{acknowledgments}

\newpage

\widetext

\section{{\Large Supplemental materials}}

\section{Derivation of the ${\bf k}\cdot{\bf p}$ Hamiltonian }

The bands near the $\Gamma$ point are derived from the $\Gamma_8$
representation. A basis for this representation consists of the four
the total angular momentum $j=3/2$ states $|3/2,m\rangle$ with
$m=3/2,1/2,-1/2,-3/2$ \cite{book}. To construct a $\bf{k}\cdot{\bf p}$
theory we need the sixteen operators that span the direct product
space of $\Gamma_8\otimes \Gamma_8$. This can be conveniently done
using powers of the $j=3/2$ operators $\check{J}_i$ which we list here
in the $\{|3/2,3/2\rangle, |3/2,1/2\rangle, |3/2,-1/2\rangle,
|3/2,-3/2\rangle \}$ basis ($\hbar=1$)
\begin{eqnarray}
\check{J}_x&=& \frac{1}{2}\left(
      \begin{array}{cccc}
        0 & \sqrt{3} & 0 & 0 \\
        \sqrt{3} & 0 & 2 & 0 \\
        0 & 2 & 0 & \sqrt{3} \\
        0 & 0 & \sqrt{3} & 0 \\
      \end{array}
    \right)\\
\check{J}_y&=& \frac{i}{2}\left(
      \begin{array}{cccc}
        0 & -\sqrt{3} & 0 & 0 \\
        \sqrt{3} & 0 & -2 & 0 \\
        0 & 2 & 0 & -\sqrt{3} \\
        0 & 0 & \sqrt{3} & 0 \\
      \end{array}
    \right)\\
\check{J}_z &=& \frac{1}{2}\left(
      \begin{array}{cccc}
        3 & 0 & 0 & 0 \\
        0 & 1 & 0 & 0 \\
        0 & 0 & -1 & 0 \\
        0 & 0 & 0 & -3 \\
      \end{array}
    \right).
\end{eqnarray}
In particular, the relevant products can be constructed using the well
known relationship for spherical symmetry $3/2\otimes 3/2 = 3 \oplus
2 \oplus 1 \oplus 0$ and then decomposing the total $j$ states into
tetrahedral representations. Using the character table for the
tetrahedral group $T_d$ given in Table~\ref{Table1}, we find:
$j=0\rightarrow A_1$, $j=1\rightarrow T_1$, $j=2\rightarrow E\oplus
T_2$, and  $j=3\rightarrow A_2\oplus T_1 \oplus T_2$. The
corresponding operators are given in Table~\ref{Table2}. In
Table~\ref{Table2}, we also give basis functions for the different
tetrahedral representations for all power of $k_i$ up to the third
power.

\begin{table}[b]
\begin{tabular}{|c|c|c|c|c|c|}
  \hline
  $REP$ & $E$ & $6C_4^2$ & $8C_3$ & $6S_4$ &  $6 \sigma_d$ \\
  \hline
  $A_1$ & 1 & 1 & 1 & 1 & 1 \\
  $A_2$ & 1 & 1 & 1 & -1 & -1  \\
  $E$ & 2 & 2 & -1 & 0 & 0  \\
  $T_1$ & 3 & -1 & 0 & 1 & -1  \\
  $T_2$ & 3 & -1 & 0 & -1 & 1  \\
  \hline
\end{tabular}
\caption{Character table for the Tetrahedral group $T_d$.}
\label{Table1}
\end{table}

\begin{table}[b]
\begin{tabular}{|c|c|c|}
  \hline
   REP & $\check{J}_i$ basis functions & $k_i$ basis functions\\
  \hline
   $A_1$ & $\check{J}_x^2+\check{J}_y^2+\check{J}_z^2 = \frac{15}{4}\check{1}_4$ &
  $k_x^2+k_y^2+k_z^2$, $k_xk_yk_z$ \\
   $A_2$ & $\check{J}_x\check{J}_y\check{J}_z$ + symmetric permutations
    & -  \\
   $E$ &
  $[(2\check{J}_z^2-\check{J}_x^2-\check{J}_y^2)/\sqrt{3},\check{J}_x^2-\check{J}_y^2]$
  & $[(2k_z^2-k_x^2-k_y^2)/\sqrt{3},k_x^2-k_y^2]$  \\
   $T_1$ & $[\check{J}_x,\check{J}_y,\check{J}_x]$,
  $[\check{J}_x^3,\check{J}_y^3,\check{J}_z^3]$  &
  $[k_x(k_y^2-k_z^2),k_y(k_z^2-k_x^2),k_z(k_x^2-k_y^2)]$ \\
    $T_2$ & $[\check{J}_y\check{J}_x\check{J}_y-\check{J}_z\check{J}_x\check{J}_z,\check{J}_z\check{J}_y\check{J}_z-\check{J}_x\check{J}_y\check{J}_x,\check{J}_y\check{J}_z\check{J}_y-\check{J}_x\check{J}_z\check{J}_x]$ & $[k_x,k_y,k_z]$, $[k_x^3,k_y^3,k_z^3]$, $[k_x(k_y^2+k_z^2),k_y(k_z^2+k_x^2),k_z(k_x^2+k_y^2)]$ \\
  &$[\check{J}_y\check{J}_z+\check{J}_z\check{J}_y,\check{J}_x\check{J}_z+\check{J}_z\check{J}_x,\check{J}_x\check{J}_y+\check{J}_y\check{J}_x]$&$[k_yk_z,k_xk_z,k_xk_y]$\\
  \hline
\end{tabular}
\caption{Tetrahedral basis functions for powers of $\check{J}_i$ and $k_i$}
\label{Table2}
\end{table}

By forming all possible invariants from these powers of $k_i$ and the
operators that span $\Gamma_8\otimes \Gamma_8$, we arrive at the ${\bf
k}\cdot{\bf p}$ Hamiltonian
\begin{eqnarray}
H&=&\alpha k^2 +\beta\sum_i k_i^2\check{J}_i^2+\gamma \sum_{i\ne
j}k_ik_j\check{J}_i\check{J}_j +
\delta \sum_i k_i(\check{J}_{i+1}\check{J}_i\check{J}_{i+1}-\check{J}_{i+2}\check{J}_i\check{J}_{i+2})\notag \\ &&
+\epsilon_1\sum_i\check{J}_ik_i(k_{i+1}^2-k_{i+2}^2)+\epsilon_2\sum_i \check{J}_i^3k_i(k_{i+1}^2-k_{i+2}^2)+\epsilon_3\sum_ik_i^3(\check{J}_{i+1}\check{J}_i\check{J}_{i+1}-\check{J}_{i+2}\check{J}_i\check{J}_{i+2})\notag \\ &&
+\epsilon_4\sum_ik_i(k_{i+1}^2+k_{i+2}^2)(\check{J}_{i+1}\check{J}_i\check{J}_{i+1}-\check{J}_{i+2}\check{J}_i\check{J}_{i+2})
\label{Ham}
\end{eqnarray}
 where $i={x,y,z}$ and $i+1=y$ if $i=x$, {\it etc}. This Hamiltonian up to second order in the $k_i$ was initially discussed by Dresselhaus \cite{dre55}.

 \section{Determination of the ${\bf k}\cdot{\bf p}$ Hamiltonian from band structures}

The band structures of YPtBi and LuPtBi, including spin-orbit coupling, were calculated using several different approximations for
exchange-correlation: (i) the standard PBE generalized gradient approximation parameterization\cite{PBE} (also referred to as ``LDA''); (ii) the
modified Becke-Johnson LDA \cite{MBJLDA} potential that was developed to yield band gaps in better agreement with experiment for a wide class of
materials;
 and the HSE06 hybrid
functional \cite{HSE06} which includes a fraction (0.25) of exact exchange. For the MBJLDA, $\alpha$=0.012 and $\beta$=1.023 as given in
Ref.~\cite{MBJLDA} were used.

In Fig.~\ref{abinitio} we show the MBJLDA and LDA/GGA results for the $\Gamma_8$ bands of LuPtBi; the corresponding plot for
YPtBi is shown in the main paper. Although there are differences in details that show up in the fitting parameters, the band
topology is similar. Figure~\ref{fig2} compares the LDA/GGA bands to hybrid functional ones over a larger energy range,
demonstrating that the band ordering is the same for all the different exchange-correlation choices with the $\Gamma_8$ states
near the chemical potential. This result is not surprising since simple tight-binding arguments (without spin-orbit) predict
that half-Heusler compounds with 18 valence electrons are (nearly zero-gap) semiconductors with the 6-fold $\Gamma_{15}$
(3-dimensional representation $\times$ 2 for spin) state at/near $E_F$. Inclusion of spin-orbit does not alter this picture:
Spin-orbit pushes the 2-fold $\Gamma_7$ state (with its downward dispersing bands) to lower energy, while the 4-fold
$\Gamma_8$ states remain near $E_F$. Thus, although there are differences in the band dispersions, the overall band topologies
are the same. In particular, {\em any} reasonable calculation of these half-Heusler materials will lead to the $j=3/2$
$\Gamma_8$ states near the chemical potential, which is the essential aspect needed for the superconductivity discussed here.

In the following we carry out detailed fits to only the MBJLDA and LDA results. The two schemes are in much better agreement
for LuPtBi than for YPtBi (shown in the main paper).  From fitting the band structures near the $\Gamma$ point we can extract
the parameters in the ${\bf k}\cdot {\bf p}$ Hamiltonian. To carry out these fits, we first extracted the parameters $\alpha$,
$\beta$, $\gamma$, and $\delta$. We found that we needed to also include the cubic term $\epsilon_1$ to correctly model the
bands in the $(1,1,0)$ direction. The other cubic parameters we set to zero.  The resultant parameters are given in
Table~\ref{Table3}. In addition, we note that the inclusion of quartic terms is required to capture an additional electron
pocket that appears near the calculated Fermi energy for some of the density functional calculations. Early experimental
results suggest that these materials are hole doped \cite{but11,bay12}, so it is likely that this electron Fermi surface does
not appear in the superconducting materials. However, we note that our main results: the appearance of nodal broken
time-reversal $d$-wave quintet pairing states and the existence of a mixed $s$-wave singlet and $p$-wave septet state with
topologically protected line nodes are not affected by this additional electron pocket.

\begin{figure}%
    \includegraphics[width=\columnwidth]{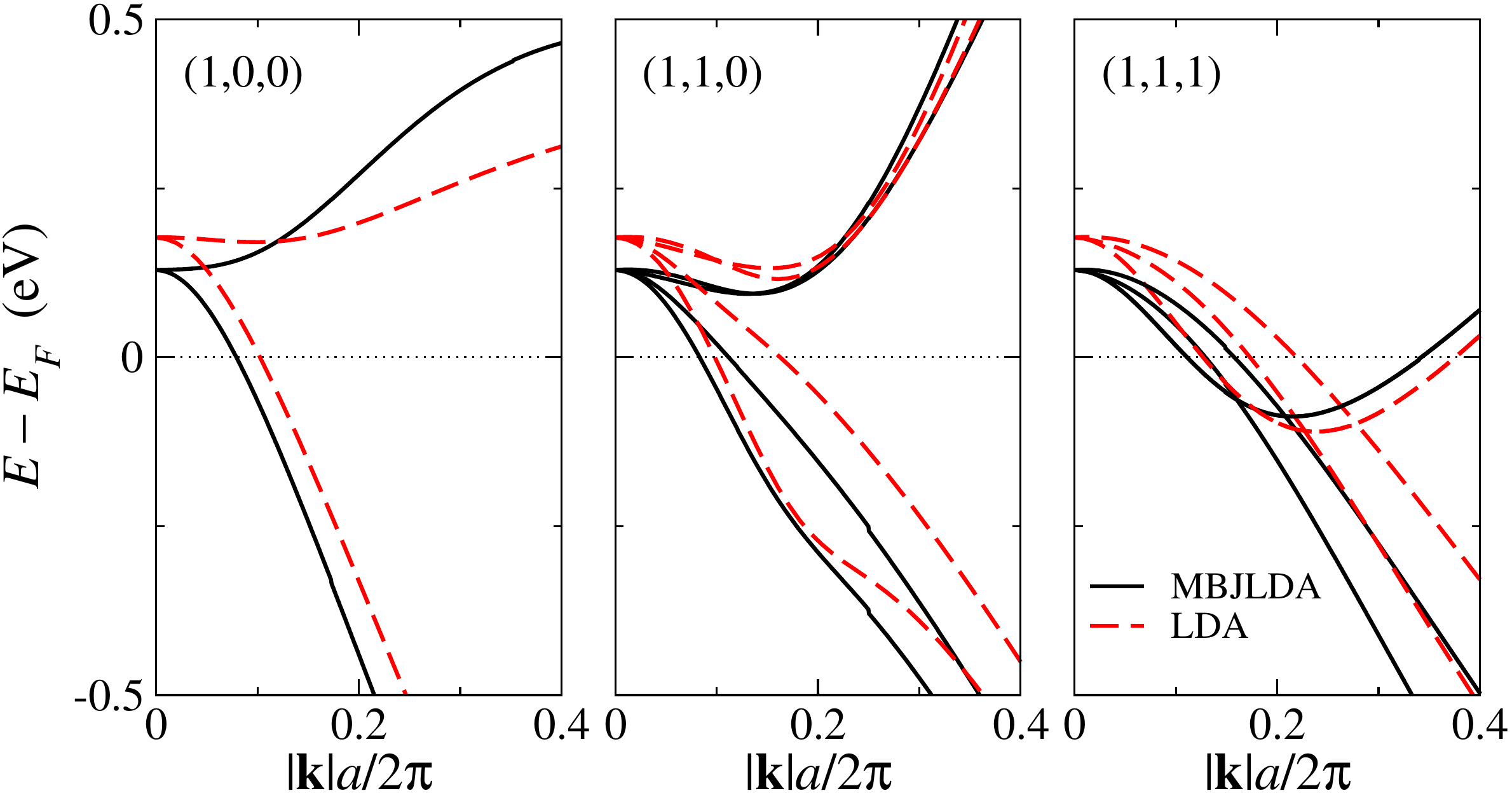}
    \caption{(Color online) Comparison of MBJLDA and LDA (PBE GGA) results for
    the $\Gamma_8$ band of LuPtBi along high symmetry directions close
    to the $\Gamma$ point. The dotted line indicates the Fermi energy
    and $a$ is the lattice constant. A similar plot for YPtBi is given in the main paper.}
    \label{abinitio}
\end{figure}

\begin{figure}%
    \includegraphics[width=\columnwidth]{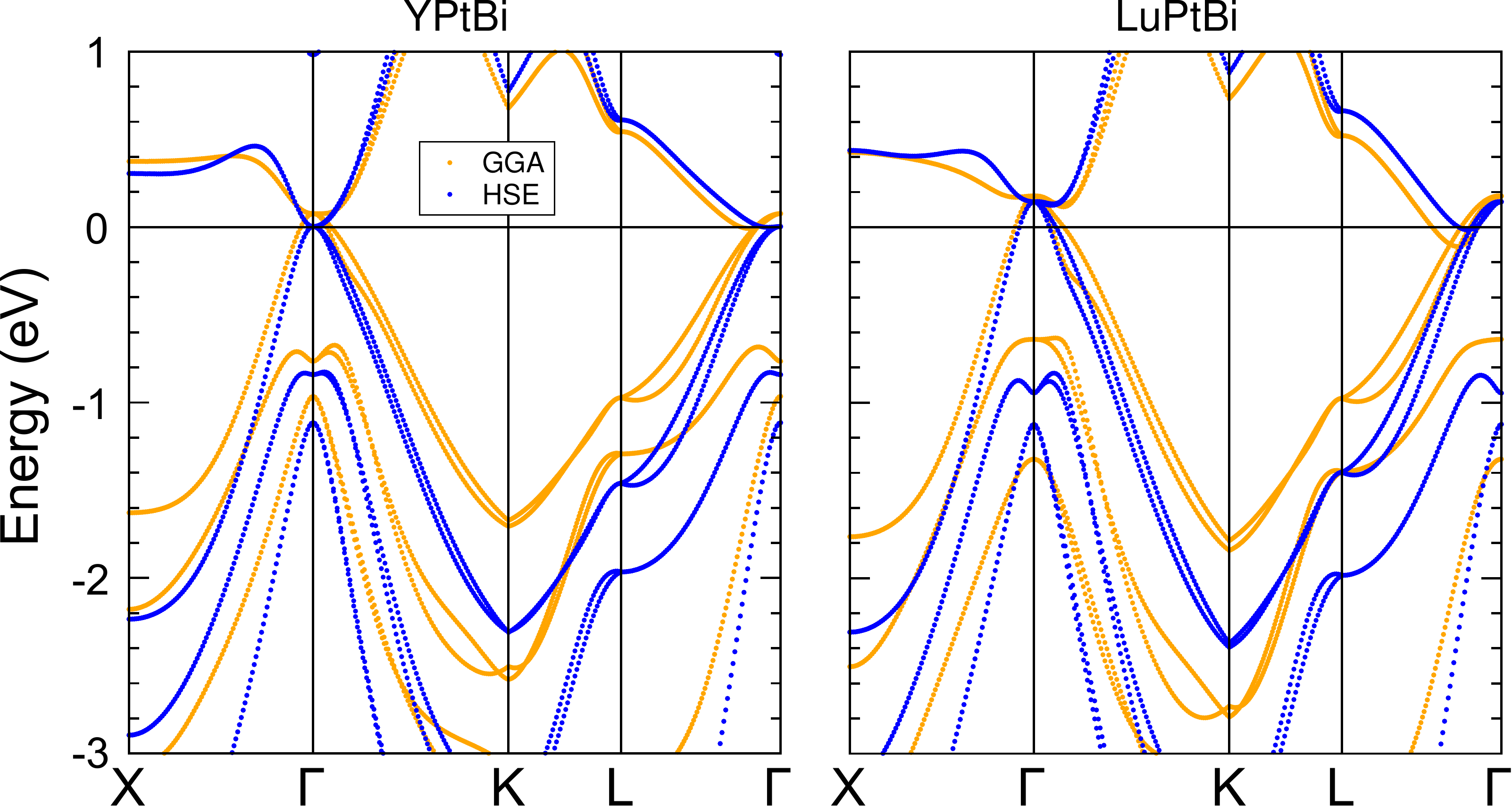}
    \caption{(Color online)  Comparison of the calculated YPtBi and LuPtBi relativistic band
structures around the Fermi level using the PBE GGA (orange) and the
HSE06 hybrid (blue) functionals. Although the hybrid functional tends to
push the valence states at the zone boundary deeper, the ordering of the
states remains the same.}
    \label{fig2}
\end{figure}

\begin{table}
\begin{tabular}{|c|c|c|c|c|c|c|}
  \hline
  Material & Potential & $\alpha$ & $\beta$ & $\gamma$ &  $\delta$ & $\epsilon_1$ \\
  &&($eVa^2/\pi^2$)&($eVa^2/\pi^2$)&($eVa^2/\pi^2$))&($eVa/\pi$)&($eVa^3/\pi^3$)\\
  \hline
  YPtBi & LDA & 8.2 & -11.6 & 1.7 & 0.01 & 73\\
  YPtBi & MBJLDA & 20.5 & -18.5  & -1.27 & 0.025& 45 \\
  LuPtBi& LDA & 0.48 &-8.7  &5.3  & 0.11 & 70\\
  LuPtBi& MBJLDA & 5.32 & -13.9 & 4.2 & 0.12& 80 \\
  \hline
\end{tabular}
\caption{${\bf k}\cdot{\bf p}$ fitting parameters for different band structures for YPtBi and LuPtBi.}
\label{Table3}
\end{table}


\begin{thebibliography}{99}

\bibitem{hhtop}S. Chadov, X. Qi, J. K\"ubler, G. H. Fecher,
C. Felser, and S. C. Zhang, Nature Mat. {\bf 9}, 541 (2010); H. Lin,
L. A. Wray, Y. Xia, S. Xu, S. Jia, R. J. Cava,
A. Bansil, and M. Z. Hasan, {\it ibid} {\bf 9}, 546
(2010); D. Xiao, Y. Yao, W. Feng, J. Wen, W. Zhu, X.-Q. Chen,
G. M. Stocks, and Z. Zhang, Phys. Rev. Lett. {\bf 105}, 096404 (2010);
W. Al-Sawai, H. Lin, R. S. Markiewicz, L. A. Wray, Y. Xia,
S.-Y. Xu, M. Z. Hasan, and A. Bansil, Phys. Rev. B {\bf 82}, 125208
(2010).
\bibitem{can91}P. C. Canfield, J. D. Thompson, W. P. Beyermann,
A. Lacerda, M. F. Hundley, E. Peterson, Z. Fisk and H. R. Ott,
J. Appl. Phys. {\bf 70}, 5800 (1991).
\bibitem{nak15}Y. Nakajima, R. Hu, K. Kirshenbaum, A. Hughes, P. Syers,
X. Wang, K. Wang, R. Wang, S. R. Saha, D. Pratt, J. W. Lynn, and
J. Paglione, Sci. Adv. {\bf 1}, e1500242 (2015).
\bibitem{gol08}G. Goll, M. Marz, A. Hamann, T. Tomanic,
K. Grube, T. Yoshino, and T. Takabatake, Physica B {\bf 403}, 1065 (2008).
\bibitem{but11} N. P. Butch, P. Syers, K. Kirshenbaum, A. P. Hope, and
J. Paglione, Phys. Rev. B {\bf 84}, 220504(R) (2011).
\bibitem{taf13} F. F. Tafti, T. Fujii, A. Juneau-Fecteau, S.
Ren\'{e} de Cotret, N. Doiron-Leyraud, A. Asamitsu, and L. Taillefer,
Phys. Rev. B {\bf 87}, 184504 (2013).
\bibitem{xu14}G. Xu, W. Wang, X. Zhang, Y. Du, E. Liu, S. Wang,
G. Wu, Z. Liu, and X. X. Zhang, Sci. Rep. {\bf 4}, 5709 (2014).
\bibitem{pan13}Y. Pan, A. M. Nikitin, T. V. Bay, Y. K. Huang,
C. Paulsen, B. H. Yan, and A. de Visser, Europhys. Lett. {\bf 104},
27001 (2013).
\bibitem{nit15}A. M. Nikitin, Y. Pan, X. Mao, R. Jehee, G. K. Araizi,
Y. K. Huang, C. Paulsen, S. C. Wu, B. H. Yan, and A. de Visser,
J. Phys.: Condens. Matter {\bf 27}, 275701 (2015).
\bibitem{bay12} T. V. Bay, T. Naka, Y. K. Huang, and A. de Visser,
Phys. Rev. B {\bf 86}, 064515 (2012).
\bibitem{jp15}H. Kim, K. Wang, Y. Nakajima, R. Hu, S. Ziemak,
P. Syers, L. Wang, H. Hodovanets, J. D. Denlinger, P. M. R. Brydon,
D. F. Agterberg, M. A. Tanatar, R. Prozorov, and J. Paglione,
arXiv:1603.03375 (unpublished, 2016).
\bibitem{ak15} A. Banerjee, A. Fang, C. Adamo, E. Levenson-Falk,
A. Kapitulnik, S. Chandra, B. Yan, and C. Felser,  March Meeting 2015
abstract T25.00005.
\bibitem{majorana}L. Mao, M. Gong, E. Dumitrescu, S. Tewari, and
C. Zhang, Phys. Rev. Lett. {\bf 108}, 177001 (2012); A. G. Moghaddam,
T. Kernreiter, M. Governale, and U. Z\"ulicke, Phys. Rev. B {\bf 89},
184507 (2014); C. Fang, B. A. Bernevig, and M. J. Gilbert,
Phys. Rev. B {\bf 91}, 165421 (2015); W. Wang, Y. Li, and C. Wu,
arXiv:1507.02768 (unpublished, 2015); A. Shitade and Y. Nagai,
arXiv:1512.07997 (unpublished, 2015).
\bibitem{ho99} T. L. Ho and S. Yip, Phys. Rev. Lett. {\bf 82}, 247 (1999).
\bibitem{suppl} See the supplemental information for the {\it ab
initio} predictions for LuPtBi, hybrid functional results, and
extension of the ${\bf k}\cdot{\bf p}$ Hamiltonian to
higher order. This includes~Refs.~\cite{book,dre55,PBE,MBJLDA,HSE06}
\bibitem{book} G.F. Koster, J.O. Dimmock, R.G. Wheeler, and H. Statz, {\it Properties of the thirty-two point groups}, (MIT Press, Cambridge, 1963).
\bibitem{dre55}G. Dresselhaus, Phys. Rev. {\bf 100}, 580 (1955).
\bibitem{PBE} J. P. Perdew, K. Burke, and M. Ernzerhof, Phys. Rev. Lett.
{\bf 77}, 3865 (1996); {\bf 78}, 1396(E) (1997).
\bibitem{MBJLDA}F. Tran and P. Blaha, Phys. Rev. Lett. {\bf 102}, 226401 (2009).
\bibitem{HSE06} J. Heyd, J. E. Peralta, G. E. Scuseria, and R. L. Martin,
J. Chem. Phys. {\bf 123}, 174101 (2005).
\bibitem{bau12}E. Bauer and M. Sigrist, editors, {\it
Non-Centrosymmetric Superconductors: Introduction and Overview},
(Springer, Heidelberg, 2012)
\bibitem{fri04}P. A. Frigeri, D. F. Agterberg, A. Koga, and M. Sigrist,
Phys. Rev. Lett. {\bf 92}, 097001 (2004).
\bibitem{sig91}M. Sigrist and K. Ueda, Rev. Mod. Phys. {\bf 63}, 243 (1991).
\bibitem{sch15}A. P. Schnyder and P. M. R. Brydon, J. Phys.:
Condens. Matter {\bf 27}, 243201 (2015).
\bibitem{nag15} Y. Nagai, Phys. Rev. B {\bf 91}, 060502(R) (2015).
\bibitem{hir88} P. J. Hirschfeld, P. W\"olfle, and D. Einzel, Phys. Rev. B
{\bf 37}, 83 (1988).
\bibitem{kon89}R. Konno and K. Ueda, Phys. Rev. B {\bf 40}, 4329
(1989).
\bibitem{mei15}M. Meinert, Phys. Rev. Lett. {\bf 116}, 137001 (2016).

\end{thebibliography}

\begin{thebibliography}{99}

\bibitem{book}  G.F. Koster, J.O. Dimmock, R.G. Wheeler, and H. Statz, {\it Properties of the thirty-two point groups}, MIT Press, Cambridge, Mass. (1963).
\bibitem{dre55}G. Dresselhaus, Phys. Rev. {\bf 100}, 580 (1955).
\bibitem{PBE}
J. P. Perdew, K. Burke, and M. Ernzerhof, Phys. Rev. Lett.
{\bf 77}, 3865 (1996); {\bf 78}, 1396 (1997).
\bibitem{MBJLDA}
F. Tran and P. Blaha, Phys. Rev. Lett. {\bf 102}, 226401 (2009).
\bibitem{HSE06}
J. Heyd, J. E. Peralta, G. E. Scuseria, and R. L. Martin, J.
Chem. Phys. {\bf 123}, 174101 (2005).
\bibitem{but11} N. P. Butch, P. Syers, K. Kirshenbaum, A. P. Hope, and
J. Paglione, Phys. Rev. B {\bf 84}, 220504(R) (2011).
\bibitem{bay12} T. V. Bay, T. Naka, Y. K. Huang, and A. de Visser,
Phys. Rev. B {\bf 86}, 064515 (2012).





\end{thebibliography}
\end{document}